\documentstyle[11pt,fleqn]{article}
\begin{document}

\title{THE ASYMPTOTIC METHOD DEVELOPED FROM WEAK TURBULENT THEORY AND THE
NONLINEAR PERMEABILITY AND DAMPING RATE IN QGP }
\author{{Z{\small HENG} X{\small IAOPING}{\hskip .5cm}
L{\small I} J{\small IARONG}}\\
{\small The Institute of Particle Physics of Huazhong Normal University}\\
{\small Wuhan,430070,P.R.China}}
\maketitle

\vskip-8truecm
\hskip12truecm HZPP-97-03

\hskip12truecm June 15, 1997
\vskip6truecm

\begin{center}
\begin{minipage}{120mm}
\vskip 3.5cm
\begin{center}{\bf ABSTRACT}\end{center}
    {\ \ \ With asymptotic method developed from weak turbulent theory,
the kinetic equations for QGP are expanded in fluctuation field
potential $A^T_\mu $. Considering the second-order and third-order currents,
we derive the nonlinear permeability tensor function from Yang-Mills field
equation, and find that the third-order current is more important in turbulent
theory. The nonlinear permeability formulae for longitudinal color oscillations
show that the non-Abelian effects are more important than the Abelian-like
effects. To compare
with other works,
we give the numerical result of the damping rate for the modes with zero wave
vector.\\
{\bf KEY WORDS}: QGP, Kinetic theory, Weak turbulent theory, Fluctuation.\\
{\bf PACS} numbers: 12.38.Mh, 51.10.+y}
\end{minipage}
\end{center}

\vfill
{\large\bf
\begin{center} INSTITUTE OF PARTICLE PHYSICS
\vskip0.2truecm

HUAZHONG NORMAL UNIVERSITY
\vskip0.2truecm

WUHAN CHINA \end{center}
}
\newpage

\indent
Kinetic theory is an important method by which it is possible to investigate
the non-linear, non-Abelian and non-equilibrium effects in quark-gluon plasma
(QGP). Some progress has been made in the study of the collective
behaviors in QGP by the applications of
classical or semiclassical kinetic equations\cite {r1, r2}. The medium
response to external fields has been preliminarily discussed. However,
the non-Abelian effects has not been demonstrated well in the works done
before, only linear effects are involved in most of those works.
Therefore, the results are similar to the Abelian case. Some
authors\cite {r3,r4} have computed the non-linear eigenfrequency shift and
non-linear Landaue damping rate with  multiple time-scale method, but the
nonlinear dielectric function, from which the properties of QGP
in non-equilibrium state
can be studied, are not discussed. In this letter, we aim at
obtaining the non-linear permeability and the collisionless damping rate
by regarding the non-linear collective effects as the results arisen from
the turbulent fluctuations in QGP.
In general, the key to calculations beyond linear approximation is
that one need to find a meaningful iterating method with which nonlinear
equation can  be effectively solved. Fortunately, a weak turbulent
theory can just help us to deal with the the problem
since the theory gives spontaneously the small quantity, the field excited
by fluctuations, which is
used to asymptotically expand the kinetic equations and Yang-Mills equation.

\indent
In weak turbulent theory, the quantities with
characteristics of collective motion can be decomposed into regular parts and
stochastic fluctuation terms\cite{r5}:
$$f=f^R+f^T,\ \ \ \ \ \bar{f}=\bar{f}^R+\bar{f}^T,\ \ \ \ \ \ \ G=G^R+G^T,$$
$$A_\mu=A^R_\mu+A^T_\mu,$$
where $f,\bar{f},G$ denote the distribution functions of quarks, antiquarks
and gluons
in QGP, respectively, $A_\mu$ is 4-vector potential, $R$ the index of the
 quantities averaged over the random phases, namely, $X^R=<X>$,
and $T$ the index of the fluctuation terms. To simplify, as is done in
electromagnetic plasma,
we let $A^R_\mu=0$, that is, the fields excited by the fluctuations of
particle density in QGP are considered, while the background fields are
ignored.

\indent
We now outline the calculations. We start from the following
equations\cite{r6,r7},
\begin{equation}
p^\mu D_\mu f+{g\over 2}p^\mu\partial^\nu_p\{F_{\mu\nu},f\}=0,
\end{equation}
\begin{equation}
p^\mu D_\mu\bar{f}-{g\over 2}p^\mu\partial^\nu_p\{F_{\mu\nu},\bar{f}\}=0,
\end{equation}
\begin{equation}
p^\mu \tilde{D}_\mu G+{g\over 2}p^\mu\partial^\nu_p\{\tilde{F}_{\mu\nu},G\}=0,
\end{equation}
\begin{equation}
D_\mu F^{\mu\nu}=j^\nu,
\end{equation}
\begin{equation}
j^\nu=-{g\over 2}\int\frac{{\rm d}p}{(2\pi)^3p_0}
      p^\nu(N_f(f-\bar{f})+2i\tau_af_{abc}G_{bc}),
\end{equation}
where $F_{\mu\nu},\tilde{F}_{\mu\nu}$ represent the field tensors in fundamental
and adjoint representations, i.e., $F_{\mu\nu}=F^a_{\mu\nu}\tau_a,
\tilde{F}_{\mu\nu}=F^a_{\mu\nu}T_a$
with  the corresponding generators $\tau_a$ and  $T_a$,
$f_{abc}$ is the structure constant of SU(3), $N_f$
the number of the flavors of quarks, $j^\nu$ the 4-current density. In weak
turbulent approximation, $A_\mu$ is small and can be used to expand
the quantities such as $f^T,\bar{f}^T $ and $G^T$.

\indent
At first, we deal with the case in linear approximation.
By averaging Eqs(1),(2)and(3)over stochastic phases, we obtain the expressions
for $f^R, \bar{f}^R$ and $G^R$,

\begin{equation}
p^{\mu}\partial_{\mu}f^{R}=-igp^{\mu}\langle[A^{T}_{\mu}, f^{T}]\rangle
-{g\over 2}p^{\mu}\langle\{F^{T}_{\mu\nu},\partial^{\nu}_{p}f^{T}\}\rangle,
\end{equation}
\begin{equation}
p^{\mu}\partial_{\mu}\bar{f}^{R}=-igp^{\mu}\langle[A^{T}_{\mu}, \bar{f}^{T}]\rangle
+{g\over 2}p^{\mu}\langle\{F^{T}_{\mu\nu},\partial^{\nu}_{p}\bar{f}^{T}\}\rangle,
\end{equation}
\begin{equation}
p^{\mu}\partial_{\mu}G^{R}=-igp^{\mu}\langle[\tilde{A}^{T}_{\mu}, G^{T}]\rangle
-{g\over 2}p^{\mu}\langle\{\tilde{F}^{T}_{\mu\nu},\partial^{\nu}_{p}G^{T}\}\rangle.
\end{equation}

Subtracting (6),(7),(8) from (1),(2),(3), respectively, we arrive at the 
equations for the fluctuation terms of the distribution functions

\begin{eqnarray}
&p^{\mu}\partial_{\mu}f^{T}+igp^{\mu}[A^{T}_{\mu}, f^{R}]+{g\over 2}p^{\mu}\partial^{\nu}_{p}\{F^{T}_{\mu\nu},\partial^{\nu}_{p}f^{R}\}\nonumber\\
&=-igp^{\mu}([A^{T}_{\mu}, f^{T}]-<[A^{T}_{\mu}, f^{T}]>)-{g\over 2}p^{\mu}\partial^{\nu}_{p}(\{F^{T}_{\mu\nu},\partial^{\nu}_{p}f^{T}\}-<\{F^{T}_{\mu\nu},\partial^{\nu}_{p}f^{T}\}>),
\end{eqnarray}

\begin{eqnarray}
&p^{\mu}\partial_{\mu}\bar{f}^{T}+igp^{\mu}[A^{T}_{\mu}, \bar{f}^{R}]- {g\over 2}p^{\mu}\partial^{\nu}_{p}\{F^{T}_{\mu\nu}, \bar{f}^{R}\}\nonumber\\
&=-igp^{\mu}([A^{T}_{\mu},\bar{f}^{T}]-<[A^{T}_{\mu}, \bar{f}^{T}]>)+{g\over 2}p^{\mu}\partial^{\nu}_{p}(\{F^{T}_{\mu\nu},\partial^{\nu}_{p}\bar{f}^{T}\}-<\{F^{T}_{\mu\nu},\partial^{\nu}_{p}\bar{f}^{T}\}>),
\end{eqnarray}

\begin{eqnarray}
&p^{\mu}\partial_{\mu}G^{T}+igp^{\mu}[\tilde{A}^{T}_{\mu}, G^{R}]+{g\over 2}p^{\mu}\partial^{\nu}_{p}\{\tilde{F}^{T}_{\mu\nu}, G^{R}\}\nonumber\\
&=-igp^{\mu}([A^{T}_{\mu}, G^{T}]-<[A^{T}_{\mu}, G^{T}]>)-{g\over 2}p^{\mu}\partial^{\nu}_{p}(\{F^{T}_{\mu\nu},\partial^{\nu}_{p}G^{T}\}-<\{F^{T}_{\mu\nu},\partial^{\nu}_{p}G^{T}\}>).
\end{eqnarray}

In linear approximation, all fluctuation quantities are kept to the first-order
in $A_\mu^T$. By using Fourier transformation, 
Eqs(9),(10)and(11) enable us to
write the fluctuation terms of the distribution functions in terms of the 
regular parts of the functions and $A_\mu^T$. Substituting the fluctuation
 terms
into Eq(5), one gives immediately the first-order current 

\begin{eqnarray}
j^{\nu(T)}&=&\int {d^3p\over (2\pi)^3p_0}p^\nu p^\rho
	   {k_\rho\delta_{\lambda\sigma }-k_\lambda\delta_{\rho\sigma }\over
	    pk+i0^+}\partial^\lambda_p(N_f(f^R+\bar f^R)+2NG^R)A^T_\sigma\nonumber\\
	  &\equiv&\Pi^{\nu\sigma}(k)A^T_\sigma.
\end{eqnarray}

Obviously, the covariant polarization tensor $\Pi^{\nu\sigma}$ should be 
determined in terms of $f^R,\bar{f}^R$ and $G^R$. Furthermore, 
the 3-dimension dielectric tensor can be
written as\cite{r8}

\begin{equation}
\epsilon_{ij}=\delta_{ij}-{1\over\omega}\Pi_{ij},
\end{equation}

and $\epsilon_{ij}$ is decomposed into longitudinal and transverse components
\cite{r8,r9,r10}

\begin{equation}
\epsilon_{ij}=\varepsilon_L(k){k_ik_j\over{\bf k}^2}+\varepsilon_T(k)
(\delta_{ij}-{k_ik_j\over{\bf k}^2}).
\end{equation}

Let us consider the ultra-relativistic case where $p_0=|{\bf p}|$.
We obtain  

\begin{equation}
\epsilon_L(\omega,{\bf k})=1+{m^2\over{\bf k}}[1+
{\omega\over 2|{\bf k}|}{\rm ln}|{\omega-|{\bf k}|\over \omega+|{\bf k}|}|
+{i\pi\omega\over 2|{\bf k}|}\Theta(|{\bf k}|-\omega)],
\end{equation}

\begin{equation}
\epsilon_T(\omega,{\bf k})=1+{m^2\over \omega |{\bf k}|}[-{\omega\over 2{\bf k}}
+{1\over 4}[1-{\omega^2\over {\bf k}^2}]{\rm ln}|{\omega-|{\bf k}|\over\omega
+|{\bf k}|}|+{i\pi\over 4}[1-{\omega^2\over {\bf k}^2}]\Theta(|{\bf k}|-\omega)],
\end{equation}
where $m^2={g\over 2\pi^2}\int dp_0p_0^2\frac{\partial}{\partial p_0}
(N_f(f^R+\bar f^R)+2NG^R)$ is the effective mass, 
and $\Theta(x)$ a step function.
If the fluctuations near equilibrium  are considered, Eqs(15) and (16) 
coincide with Ref\cite {r9} completely, and $f^R, \bar f^R$  are
determined by Feimi-Dirac distribution $n_F$, $G^R$ 
by Bose-Einstein distribution $n_B$, and 
$m=m_D=gT\sqrt{N+N_f/2\over 3}$ which is the Debye screening mass,
as determined from
a 1-loop approximation in thermal QCD\cite{r11}. If we
discuss the non-equilibrium QGP, the mass $m$ will deviate from $m_D$ 
because
of the difference of the distribution functions from those in equilibrium.

\indent
In linear approximation, a possible damping mechanism for those collective
modes arises from collisions of the plasma wave with individual particles.
But one can check that the Landau damping has never happened as is done 
in Ref\cite{r1,r6,r7}.

\indent
We now turn to study the dielectric properties beyond the linear approximation. 
In this case, the high-order kinetic equations should be involved. 
Here only the second-order and the third-order equations are considered. With 
recurrent iterating method in solving the kinetic equations, the second-
order and third-order fluctuation terms are  given in terms of the square 
and cube of $A_\mu^T$, $f^R $, $\bar f^R$ and $G^R$. Furthermore, the 
second-order and third-order currents can also be determined. 
According to the field equation (4), 
one finds easily in the case of longitudinal oscillations that

\begin{eqnarray}
&-&{1\over 2}\epsilon^l({\bf k})<dg>\nonumber\\
&=&g\int dk_1dk_2\delta(k-k_1-k_2){{\bf(kk_1)
(kk_2)\over |k||k_1|k_2|}}tr(\tau_d[\tau_a,\tau_b])<abg>\nonumber\\
&-&g^2\int dk_1dk_2dk_3\delta(k-k_1-k_2-k_3){{\bf(k_1k_2)
(kk_3)\over |k||k_1|k_2||k_3|}}tr(\tau_d[\tau_a,[\tau_b,\tau_c]])<abcg>\nonumber\\
&+&{gm^2\over 4}\int dk_1dk_2\delta(k-k_1-k_2)tr(\tau_d[\tau_a,\tau_b])
<abg>\omega_1W_{k,k_1,k_2}\nonumber\\
&-&{g^2m^2\over 4}\int dk_1dk_2dk_3\delta(k-k_1-k_2-k_3)tr(\tau_d[\tau_a,[\tau_b,\tau_c]])\nonumber\\
&\times &(<abcg>-<ag><bc>)\omega_1\omega_3\Gamma_{k,k_1,k_2,k_3}\nonumber\\
&-&{g^2m'^2\over 16}\int dk_1dk_2dk_3\delta(k-k_1-k_2-k_3)tr(\tau_d\{\tau_a,\{\tau_b,\tau_c\}\})\nonumber\\
&\times &(<abcg>-<ag><bc>)\omega_1\omega_2\omega_3\Gamma_{k,k_1,k_2,k_3}\nonumber\\
&-&{g^2m'^2_g\over 4}\int dk_1dk_2dk_3\delta (k-k_1-k_2-k_3)(\delta_{bc}\delta_{ad}
+\delta_{ab}\delta_{cd}+\delta_{ac}\delta_{bd})\nonumber\\
&\times &(<abcg>-<ag><bc>)\omega_1\omega_2\omega_3\Gamma_{k,k_1,k_2,k_3},
\end{eqnarray}

with $$m'^2={g\over 2\pi^2}\int p_0^2dp_0\frac{\partial^3}{\partial p_0^3}
      (N_f(f^R+\bar f^R)+2NG^R),$$
     $$m'^2_g={g\over 2\pi^2}\int p_0^2dp_0\frac{\partial^3}{\partial p_0^3}
      G^R,$$
     $$W_{k,k_1,k_2}=\int {d\Omega\over 4\pi}{\bf vk\over |k|}{\bf vk_1\over |k_1|}
      {\bf vk_2\over |k_2|}{1\over vk+i0^+}{1\over vk_2+i0^+},$$
     $$\Gamma_{k,k_1,k_2,k_3}=\int {d\Omega\over 4\pi}{\bf vk\over |k|}{\bf vk_1\over |k_1|}
      {\bf vk_2\over |k_2|}{\bf vk_3\over |k_3|}{1\over vk+i0^+}{1\over
      v(k_2+k_3)+i0^+}{1\over vk_3+i0^+},$$
where $\epsilon^l(k) $ denotes the linear permeability  for longitudinal modes, 
i.e., $\epsilon_L(k)$ mentioned above, 
${\bf v}={\bf p\over |p|}$ in $v^\mu\equiv (1,{\bf v})$
is the velocity of the particles in QGP, $\int d\Omega$ runs over all the 
directions of the unit vector ${\bf v}$. For the sake of simplicity,
we have already used the following marks in Eq(17) 
$$<ab>=<A^T_a(k_1)A^T_b(k_2)>,$$
$$<abc>=<A^T_a(k_1)A^T_b(k_2)A^T_c(k_3)>,$$
$$<abcd>=<A^T_a(k_1)A^T_b(k_2)A^T_c(k_3)A^T_d(k_4)>,$$
where $< >$ represents averaging over stochastic phases. The average values above
are called  two-point, three-point and four-point correlations, 
respectively. We can approximatively split the four-point correlations into
three pairwise products of two-point correlations in weak turbulent theory,
whereas this procedure gives zero for the three-point correlations since we know 
$<a>=<A^T_a(k)>=0$. These can be expressed as
$$<abcd>\approx <ab><cd>+<ac><bd>+<ad><bc>$$
and
$$<abc>\approx <a><b><c>+<ab><c>+<bc><a>+<ac><b>=0.$$
Here we have taken the two-point correlations
as $<ab>\equiv <A^T_a(k_1)A^T_b(k_1)>\delta (k_1+k_2)$ for a stationary 
uniform turbulent fluctuation\cite{r5}. 
Marking $<A^T_a(k)A^T_b(k)>\equiv <ab>$, Eq(17) then can be written as
\begin{eqnarray}
&+&[{1\over 2}\omega^2\epsilon^l\delta_{df}+{1\over 2}g^2f_{bfe}f_{ade}\int
dk_1<ab> [1+gm^2(\omega\Gamma_{k,k_1,-k_1,k}+\omega_1\Gamma_{k,k_1,k,-k_1})]\nonumber\\
&-&{1\over 4}g^3\int dk_1<ab>[m'^2({1\over N}\delta_{ad}\delta_{bf}+{1\over 2}
d_{bfe}d_{ade})+4m'^2_g(\delta_{ad}\delta_{bf}+\delta_{ab}\delta_{df}
+\delta_{af}\delta_{bd})]\nonumber\\
&\times &\omega\omega_1 (\Gamma_{k,k_1,-k_1,k}+\Gamma_{k,k_1,k,-k_1})]\bullet <fg>=0,
\end{eqnarray}
where $d_{abc}$ is the symmetric constant in the generators algebra of SU(3).
Since $<fg>\not= 0$, its coefficient in Eq(18) should
vanish. It is the dispersion equation which contains the contributions of
non-linear currents.
The total permeability is the sum of two terms
$$\epsilon_{df}(k)=\epsilon^l\delta_{df}+\epsilon^N_{df},$$
here $\epsilon^N_{df}$ is the nonlinear permeability as defined by

\begin{eqnarray}
\epsilon_{df}^N&=&{g^2\over \omega^2}\int dk_1<ab>\nonumber\\
&\times & [f_{abe}f_{fde}+{1\over 2}g^2m_D^2f_{bfe}f_{ade}
(\omega\Gamma_{k,k_1,-k_1,k}+\omega_1\Gamma_{k,k_1,k,-k_1})\nonumber\\
&-&{1\over 8}g^2m'^2({1\over N}\delta_{ad}\delta_{bf}+{1\over 2}d_{bfe}d_{ade})
\omega\omega '^2(\Gamma_{k,k_1,-k_1,k}+\Gamma_{k,k_1,k,-k_1})\nonumber\\
&-&{1\over 4}g^2m_g^2(\delta_{bf}\delta_{ad}+\delta_{ab}\delta_{df}
+\delta_{af}\delta_{bd})\omega\omega^2_1(\Gamma_{k,k_1,-k_1,k}+\Gamma_{k,k_1,k,-k_1})].
\end{eqnarray}

The above formula shows that the non-linear permeability in QGP is an $8\times 8$
matrix in color space and contains the structure constant of SU(3), 
which implies the that the non-Abelian features have been reflected in
dispersion relations. 
One sees that it is very difficult
to analytically integrate (19) because of its complicated structure, 
the numerical 
technique is required.  As an example, we only compute the eigenmodes 
with the same color for the equilibrium QGP with 
random thermal fluctuations,
$$<ab>=<(A^T(k))^2>\delta_{ab},$$
$$<(A^T(k))^2>={\pi\over\omega^2}[\delta(\omega -\omega ({\bf k}))
+\delta(\omega -\omega({\bf k}))]I({\bf k}),$$
where $I({\bf k})$ characterizes the total intensity of the
fluctuation oscillations with frequencies $\omega ({\bf k})$ and $-\omega 
({\bf k})$.  Therefore the nonlinear permeability matrix becomes a scalar
coefficient

\begin{eqnarray}
\epsilon^N&&=-{Ng^2m_D^2\over 4\omega^2}\int dk_1I({\bf k_1})
	      {\pi\over\omega_1^2}\delta(\omega_1-\omega_1({\bf k_1}))
	      (\omega\Gamma_{k,k_1,-k_1,k}+\omega_1\Gamma_{k,k_1,k,-k_1})\nonumber\\
	   &&+{N_f(N^2-4)\over 16N\pi^2\omega^2}g^4\int dk_1I({\bf k_1})
	      {\pi\over\omega_1^2}\delta(\omega_1-\omega_1({\bf k_1}))
	      \omega\omega_1^2(\Gamma_{k,k_1,-k_1,k}+\Gamma_{k,k_1,k,-k_1})\nonumber\\
	   &&+{5N^2\over 8\pi^2\omega^2}g^4({3\over g}+{5\over 2})\int dk_1I({\bf k_1})
	      {\pi\over\omega_1^2}\delta(\omega_1-\omega_1({\bf k_1}))
	      \omega\omega_1^2(\Gamma_{k,k_1,-k_1,k}+\Gamma_{k,k_1,k,-k_1}).\nonumber\\
	   && \ \ \ \ \ \ \ \ \ \ \ \ \ \ \ \ \ \ \ \ \ \ \ \ \ \ \ \ \ \ \ \ \ \ \ \ \  \ \ \ \ \ \ \ \
\ \ \ \ 
\end{eqnarray}

The integrals in Eq(20) contain three complex factors denoted by 
${1\over vk+i0^+}, {1\over vk_1+i0^+}$ and ${1\over v(k-k_1)+i0^+}$.
According to the Cauchy principal value formula, three delta functions 
$\delta (vk), \delta (vk_1)$ and $\delta (v(k+k_1))$ should emerge
in the imaginary parts of the integrals. However, one sees
that the classical eigenmodes in QGP are timelike \cite{r1} from the linear-
response properties, so that the Chevenkov's condition ($vk=0$) can not
be satisfied at all. 
Therefore, only the last factor arising from nonlinear processes makes 
contribution to the imaginary parts of 
the integrals. Since one can check that the condition for the nonvanishing
of the imaginary parts mentioned above is the existence of the delta function 
$\delta (\omega-\omega_1-{\bf v(k-k_1)})$, the upper-bound of the integrals
over ${\bf k_1}$ in Eq(20) should equal to $gT$ approximately in the long-wavelength
region. In this region, the dispersion relation in
linear approximation is reduced to the form of\cite{r10,r11}
$$\omega ({\bf k})=\omega^2_p+{3\over 5}{\bf k}^2,$$
and one can obtain from random thermal fluctuations\cite {r4}
$$I({\bf k})=4\pi T,$$
where $T$  is the temperature of equilibrium QGP. After these
considerations, one can estimate that the first term in Eq(20) representing
pure non-Abelian effects is of order $g$, while the last two terms 
(Note: there are similar terms in a electric-magnetic plasma) are of
orders $g^2$ and $g^3$, respectively. This implies that the pure 
non-Abelian effect is more important in the nonlinear response theory. 
To compare with other works\cite{r12,r13}, as an example, we prepare to 
work out the result for ${\bf k}=0$ modes.
Calculating of $\epsilon^N$ in Eq(20) to the order in $g$, we have

\begin{eqnarray}
\epsilon^N(\omega ,{\bf k})&=&{Ng^2m_D^2\over 16\pi^2\omega^3}\int d\omega_1 d|{\bf k_1}|
I({\bf k_1}){{\bf k}^2_1\over \omega_1^2}\delta (\omega_1-\omega_1({\bf k_1}))\nonumber\\
			   &\times & [{2\over \omega_1}
(1-{\omega^2_1\over 2\omega |{\bf k_1}|}{\rm ln}|{1+{\omega_1\over |{\bf k_1}|}
\over 1-{\omega_1\over |{\bf k_1}|}}|)\nonumber\\
			    &+& {2(\omega_1-\omega)\over \omega_1^2}
(1+{(\omega_1-\omega)^2\over 2\omega |{\bf k_1}|}{\rm ln}|{1+{\omega_1-\omega
\over |{\bf k_1}|}
\over 1-{\omega_1-\omega\over |{\bf k_1}|}}|)\nonumber\\
			    &-&{i\pi(\omega_1-\omega)^3\over \omega\omega_1^2|{\bf k_1}|}
\Theta (1-{\omega_1-\omega\over |{\bf k_1}|})].
\end{eqnarray}
We immediately obtain the numerical result of the imaginary part of $\epsilon^N$
\begin {equation}
{\rm Im}\epsilon^N\approx -0.72g.
\end{equation}
Then the damping rate
\begin{eqnarray}
\gamma^N&=&-{{\rm Im}\epsilon^N\over {\partial\epsilon^l\over \partial\omega}}\nonumber\\
	&=&0.25g^2T.
\end{eqnarray}
For pure gluon gas, we have
\begin{equation}
\gamma^N=0.20g^2T.
\end{equation}

Eq(24) coincides with the results based on the resummation of the hard thermal 
loops\cite{r12} and the asymptotic calculations with the multiple time-scale
method\cite{r13}. In fact, more situations
can be discussed from our method as long as one
knows the knowledge of the mean distribution functions denoted by index
$R$ in a nonequilibrium plasma.

\indent
Finally, we summarize our paper. 

\indent
(1) It is an effective method to apply the weak turbulent theory to the 
study of nonlinear problems in QGP.
In principle, we can derive any permeability and damping rate (collisionless)
beyond linear approximation. In this paper, We compute only the 
effects coming from the second-order and third-order currents 
in the processes of 
three or four waves. The formulae of 
the nonlinear permeability and damping rate are obtained. For ${\bf k}=0$ modes,
the numerical result of the damping rate coincides with 
the results obtained from other methods. Our results have already show that
the contributions of pure non-Abelian 
terms to  nonlinear effects are more important than those of Abelian-like terms.
Although we know that the resonant condition of 
eigenwave with individual particle, i.e., $vk=0$,
is not satisfied, the resonance of the particle with the secondary wave coming
from coupling of eigenwaves in nonlinear processes is possible, i. e. 
the  condition, $v\Delta k=0$, can be satisfied. 

\indent
(2) Three wave correlations make no contribution to the nonlinear effects
in our results because of the special treatment in this paper. In fact three
wave processes must be more carefully expanded with four wave process\cite{r5}.
This problem will be discussed in detail in another paper.

\indent
(3) It is emphasized that weak turbulent theory not only provides 
a method 
to study the nonlinear problems in QGP, but also can advance such studies
into an important and specific physics field.

\indent
This work is supported by the National Nature Science Fund of China.

\end{document}